\documentstyle[12pt]{article}
\begin{document}
\centerline{\Large\bf Lagrangian and Hamiltonian Formulation of}
\centerline{\Large\bf Classical Electrodynamics without Potentials}
\vskip .7in
\centerline{Dan N. Vollick}
\vskip .2in
\centerline{Irving K. Barber School of Arts and Sciences}
\centerline{University of British Columbia Okanagan}
\centerline{3333 University Way}
\centerline{Kelowna, B.C.}
\centerline{Canada}
\centerline{V1V 1V7}
\vskip 0.5in
\centerline{\bf\large Abstract}
\vskip 0.5in
\noindent
In the standard Lagrangian and Hamiltonian approach to Maxwell's theory the potentials $A^{\mu}$ are taken as the dynamical variables. In this paper I take the electric field $\vec{E}$ and the magnetic field $\vec{B}$ as the the dynamical variables. I find a Lagrangian that gives the dynamical Maxwell equations and include the constraint equations by using Lagrange multipliers. In passing to the Hamiltonian
one finds that the canonical momenta $\vec{\Pi}_E$ and $\vec{\Pi}_B$ are constrained giving 6 second class constraints at each point in space. Gauss's law and $\vec{\nabla}\cdot\vec{B}=0$
can than be added in as additional constraints. There are now 8 second class constraints, leaving 4 phase space degrees of freedom. The Dirac bracket is then introduced and is calculated for the field variables and their conjugate momenta.
\newpage
\section{Introduction}
In the standard Lagrangian and Hamiltonian approach to Maxwell's theory the potentials $A^{\mu}$ are taken as the dynamical variables (see for example \cite{Jac1, Lan1}). The Lagrangian $L=\frac{1}{4}F^{\mu\nu}F_{\mu\nu}$, where $F_{\mu\nu}=\partial_{\mu}A_{\nu}-\partial_{\nu}A_{\mu}$, gives Gauss's law and Maxwell's version of Ampere's law without sources while,
$F_{[\mu\nu,\alpha]}=0$ gives Faraday's law and $\vec{\nabla}\cdot\vec{B}=0$. The canonical momenta are given by $\Pi^{\mu}=F^{\mu 0}$, so that one runs into a problem in the Hamiltonian approach since $\Pi^0=0$. This problem can be dealt with by using Dirac's approach \cite{Dir1,Dir2} for systems with constraints. The consistency of the primary constraint $\Pi^0=0$ implies a secondary constraint
$\partial_k\Pi^k=0$. These two constraints are first class and together generate a set of transformations that includes the standard gauge transformations.

In this paper I find a Lagrangian with $\vec{E}$ and $\vec{B}$ as the dynamical variables that gives the dynamical Maxwell equations. The constraint equations can be included by adding them to this Lagrangian using Lagrange multipliers. The canonical momenta $\vec{\Pi}_E$ and $\vec{\Pi}_B$ are constrained giving 6 second class constraints at each point in space. Gauss's law and $\vec{\nabla}\cdot\vec{B}=0$
can than be added in as additional constraints. There are now 8 second class constraints at each point in space leaving 4 phase space degrees of freedom. The Dirac bracket is then introduced and is calculated for the field variables and their conjugate momenta.
\section{The Lagrangian Formulation}
Maxwell's equations can be divided into the dynamical equations (with $c=1$)
\begin{equation}
\vec{\nabla}\times\vec{E}=-\frac{\partial\vec{B}}{\partial t}\hskip 1.0 in
\vec{\nabla}\times\vec{B}=\frac{\partial\vec{E}}{\partial t}+\frac{4\pi}{c}\vec{j}
\label{dyn}
\end{equation}
and the constraint equations
\begin{equation}
\vec{\nabla}\cdot\vec{E}=4\pi\rho \hskip 1.0 in \vec{\nabla}\cdot\vec{B}=0.
\label{const}
\end{equation}
In this section I will find a Lagrangian that involves $\vec{E}$ and $\vec{B}$ and their first derivatives that gives the dynamical equations of motion. The constraint equations can then be included using Lagrange multipliers.

Consider the Lagrangian
\begin{equation}
L_D=a\vec{E}\cdot\left(\frac{\partial\vec{B}}{\partial t}\right)+b\vec{B}\cdot\left(\frac{\partial\vec{E}}{\partial t}\right)+c\vec{E}\cdot\left(\vec{\nabla}\times \vec{B}\right)+d\vec{E}\cdot\left(\vec{\nabla}\times \vec{E}\right)+e\vec{B}\cdot\left(\vec{\nabla}\times \vec{B}\right),
\end{equation}
where $a,b,c,d$ and $e$ are constants. I have not included a term of the form $\vec{B}\cdot\left(\vec{\nabla}\times \vec{E}\right)$
since $\vec{B}\cdot\left(\vec{\nabla}\times \vec{E}\right)=\vec{E}\cdot\left(\vec{\nabla}\times \vec{B}\right)+\vec{\nabla}\cdot(\vec{E}\times\vec{B})$.
The Euler-Lagrange equations ($\mu=0,1,2,3$ and $k=1,2,3$)
\begin{equation}
\frac{\partial}{\partial x^{\mu}}\left(\frac{\partial L_D}{\partial(\partial_{\mu}E_k)}\right)-\frac{\partial L_D}{\partial E_k}=0
\end{equation}
give
\begin{equation}
(b-a)\frac{\partial\vec{B}}{\partial t}-c(\vec{\nabla}\times \vec{B})-2d(\vec{\nabla}\times \vec{E})=0\;.
\end{equation}
To obtain Faraday's law we require that
\begin{equation}
c=0 \hskip 0.5in and \hskip 0.5in a-b=2d\;,
\end{equation}
with $a-b\neq 0$. The Euler-Lagrange equations
\begin{equation}
\frac{\partial}{\partial x^{\mu}}\left(\frac{\partial L_D}{\partial(\partial_{\mu}B_k)}\right)-\frac{\partial L_D}{\partial B_k}=0
\end{equation}
give
\begin{equation}
(a-b)\frac{\partial\vec{E}}{\partial t}-2e(\vec{\nabla}\times \vec{B})=0\;.
\end{equation}
To obtain Ampere's law without sources we require that
\begin{equation}
a-b=2e\;.
\end{equation}
From this we can also conclude that $d=e=\frac{1}{2}(a-b)$. Thus, the Lagrangian
\begin{equation}
L_D=f\left[\vec{E}\cdot\left(\vec{\nabla}\times\vec{E}+\frac{\partial\vec{B}}{\partial t}\right)+
\vec{B}\cdot\left(\vec{\nabla}\times\vec{B}-\frac{\partial\vec{E}}{\partial t}\right)\right]
\label{Lag1}
\end{equation}
gives the dynamical Maxwell equations for arbitrary $f=\frac{1}{2}(a-b)\neq 0$. I also used the relationship $\vec{E}\cdot\frac{\partial\vec{B}}{\partial t}=-\vec{B}\cdot\frac{\partial\vec{E}}{\partial t}+\frac{\partial(\vec{E}\cdot\vec{B})}{\partial t}$ to derive (\ref{Lag1}). Throughout the rest of the paper I will take $f=1$.

Sources can be included by adding
\begin{equation}
L_S=-8\pi\vec{j}\cdot\vec{B}
\label{sources}
\end{equation}
to $L_D$.

Now consider the constraint equations. One approach is to add them to to the dynamical equations outside the action principle. Another approach is to include them in the action with the use of Lagrange multipliers. However, as we shall see below, additional constraints have to be imposed on the Lagrange multipliers to obtain Maxwell's theory. In this approach
\begin{equation}
L_C=\lambda_1\left(\vec{\nabla}\cdot\vec{E}-4\pi\rho\right)+\lambda_2\left(\vec{\nabla}\cdot\vec{B}\right)
\end{equation}
is added to $L_D+L_S$, where $\lambda_1(\vec{x},t)$ and $\lambda_2(\vec{x},t)$ are Lagrange multipliers (or auxiliary fields).
The addition of this term modifies the dynamical equations of motion. They become
\begin{equation}
\vec{\nabla}\times\vec{E}=-\frac{\partial\vec{B}}{\partial t}+\vec{\nabla}\lambda_1
\label{L1}
\end{equation}
and
\begin{equation}
\vec{\nabla}\times\vec{B}=\frac{\partial\vec{E}}{\partial t}+4\pi\vec{j}+\vec{\nabla}\lambda_2\;.
\label{L2}
\end{equation}
The additional terms involving $\lambda_1$ and $\lambda_2$ act as electric and magnetic current sources for the fields. This approach yields Maxwell's equations if and only if these sources vanish.
Taking the divergence of both sides of these equations, and assuming charge conservation, gives
\begin{equation}
\nabla^2\lambda_1=\nabla^2\lambda_2=0\;.
\end{equation}
These equations do not fix $\lambda_1$ or $\lambda_2$, so some additional conditions are required. We are free to choose conditions that set $\vec{\nabla}\lambda_1=\vec{\nabla}\lambda_2=0$, so that the equations of motion are Maxwell's equations.
One possibility is to require that
$\lambda_1$ and $\lambda_2$ vanish uniformly as $r\rightarrow \infty$. This implies that $\lambda_1=\lambda_2=0$. Another possibility is to require that $\lambda_1$ and $\lambda_2$ are bounded everywhere. According to Liouville's theorem any harmonic function $\chi(\vec{x})$ that is bounded on all of $R^3$ must be a constant. Since $\lambda_1$ and $\lambda_2$ are functions of $\vec{x}$ and $t$ we will have $\lambda_1=C_1(t)$ and $\lambda_2=C_2(t)$ implying that $\vec{\nabla}\lambda_1=\vec{\nabla}\lambda_2=0$.

The properties of the Lagrangian and constraints under Lorentz transformations are discussed in Appendix I. It is shown that the Lagrangian and constraints can be written as
\begin{equation}
L_{\alpha}=\alpha^{\mu}\left[F_{\mu\nu}\partial_{\beta}G^{\nu\beta}-G_{\mu\nu}\partial_{\beta}F^{\nu\beta}
+8\pi G_{\mu\beta}j^{\beta}\right]\;,
\end{equation}
\begin{equation}
\alpha^{\mu}[\partial_{\alpha}F^{\alpha}_{\;\;\mu}+j_\mu]=0\;\;\;\;\;\;\;\;\;\;\; and \;\;\;\;\;\;\;\;\;\;\;
\alpha^{\mu}[\partial_{\alpha}G^{\alpha}_{\;\;\mu}]=0\;,
\end{equation}
where $F_{\mu\nu}$ is the electromagnetic field strength tensor, $G^{\mu\nu}=\frac{1}{2}\epsilon^{\mu\nu\alpha\beta}F_{\alpha\beta}$
is its dual, $j^{\mu}$ is the current density and $\alpha^{\mu}=(1,\vec{0})$. It is also shown that $L_{\alpha}$ and the constraints give Maxwell's equations in all inertial frames for any constant time-like four vector $\alpha^{\mu}$. This gives a manifestly covariant formulation of the theory.

There have been previous papers that have discussed electrodynamics without potentials. The approaches taken in these papers differ from the approach taken here.
In \cite{Inf1} Infeld and Plebanski formulate a Lagrangian approach to classical electrodynamics without the use of potentials. Their action is
\begin{equation}
S=\int[\Omega_0(F)+A_{\alpha}F^{\alpha\beta}_{\;\;\;\;\;\; ;\beta}]\sqrt{g}d^4x
\end{equation}
where $F=-\frac{1}{4}F^{\alpha\beta}F_{\alpha\beta}$ and the $A_{\alpha}$ are Lagrange multipliers. The field equations that follow from this action are
\begin{equation}
F^{\alpha\beta}_{\;\;\;\;\;\; ;\beta}=0
\end{equation}
and
\begin{equation}
\Omega_{0F}F_{\alpha\beta}=A_{\beta,\alpha}-A_{\alpha,\beta}\;,
\label{BI}
\end{equation}
where $\Omega_{0F}=\frac{d \Omega_0}{dF}$.
These are the field equations equations for non-linear electrodynamics introduced by Born \cite{Bor1} and by Born and Infeld \cite{Bor2}. Maxwell's equations are obtained when $\Omega_0=F$. This approach is quite different from the one taken here. In the approach taken by Infeld and Plebanski the field equations $\partial_{\alpha}F^{\alpha\mu}=0$, which contain one constraint equation and and one (vector) dynamical equation, are enforced by Lagrange multipliers. Also, from (\ref{BI}) it can be seen that the Lagrange multiplier $A_{\alpha}$ is really the vector potential. In the approach taken here only the constraint equations are enforced by Lagrange multipliers and there is no quantity which plays the role of the vector potential. The Hamiltonian formalism is also not investigated in \cite{Inf1}.

In \cite{Fiu1}  Fiutak and Zukowski take the action to be (in their notation)
\begin{equation}
W=-\frac{1}{4}\int f_{\mu\nu}f^{\mu\nu} d^4x
\end{equation}
where $f_{\mu\nu}$ is required to satisfy the ``equations of constraint"
\begin{equation}
\partial_{\alpha}f_{\beta\gamma}+\partial_{\gamma}f_{\alpha\beta}+\partial_{\beta}f_{\gamma\alpha}=0\;.
\end{equation}
They are, therefore, assuming half of Maxwell's equations at this point (one constraint equation and and one vector dynamical equation).
The action is varied with respect to $f_{\mu\nu}$ giving
\begin{equation}
\delta_0 W=-\frac{1}{2}\int f^{\mu\nu}\delta_0f_{\mu\nu} d^4x=0
\end{equation}
The $\delta_0f_{\mu\nu}$ are not independent but must satisfy
\begin{equation}
\partial_{\alpha}\delta_0f_{\beta\gamma}+\partial_{\gamma}\delta_0f_{\alpha\beta}+\partial_{\beta}\delta_0f_{\gamma\alpha}=0\;.
\end{equation}
These constraints are taken into account by taking $\delta_0f_{\mu\nu}$ to be of the form
\begin{equation}
\delta_0f_{\mu\nu}=\partial_{\mu}\psi_{\nu}-\partial_{\nu}\psi_{\nu}
\label{variation}
\end{equation}
Fiutak and Zukowski then state ``Note that all we need is the form of the variation (\ref{variation}), and this implies that the electromagnetic potentials are not introduced." (I have changed the equation number to match the number here). This approach is also mentioned in \cite{Ogi1}.

In \cite{Str1} Strazhev and Shkol'nikov develop classical and quantum theories of the electromagnetic field without potentials. They start with the action (in their notation)
\begin{equation}
W=-\frac{1}{4}\int\Phi_{\mu\nu}^2d^4x\;\;\;\;\;\;\;\;\;\;\; where \;\;\;\;\;\;\;\;\;\;\;
\Phi_{\mu\nu}=\partial_{[\mu}\int F_{|\rho|\nu]}f_{\rho}(x-x')dx'\;,
\end{equation}
\begin{equation}
f_{\mu}(x)=\frac{1}{2}\int_0^{\infty}[\delta(x-\xi)-\delta(x+\xi)]d\xi_{\mu}
\end{equation}
and $\xi_{\mu}=\xi_{\mu}(\eta)$ is a spacelike path with $\xi_{\mu}(0)=0$ and $\xi_{\mu}(\eta)\rightarrow +\infty$ as $\eta\rightarrow +\infty$. The dynamical variables are taken to be $F_{\mu\nu}(x)$ and they are taken to satisfy $\partial_{\nu}\tilde{F}_{\mu\nu}=0$ where $\tilde{F}_{\mu\nu}$ is the dual of $F_{\mu\nu}$. They have, therefore, assumed half of Maxwell's equations by imposing this condition on $\tilde{F}_{\mu\nu}$. They then add $\int\lambda_{\mu}\partial_{\nu}\tilde{F}_{\mu\nu}d^4x$ to the action and state ``In order to avoid misunderstandings we stress that the Lagrange multipliers $\lambda_{\mu}$ are not
varied, since the relations (5) are, by definition, known independently of the variational procedure.
Hence the $\lambda_{\mu}$ cannot be treated as auxiliary dynamical variables." Here (5) is the equation $\partial_{\nu}\tilde{F}_{\mu\nu}=0$. The variation of the action with respect to $F_{\mu\nu}$ using $\partial_{\nu}\tilde{F}_{\mu\nu}=0$ does not give the remaining Maxwell equations but instead gives
\begin{equation}
\int dx'[\partial^{'}_{\nu}F_{\alpha\nu}(x')]f_{\beta}(x-x')-\partial^{'}_{\nu}F_{\beta\nu}(x')f_{\alpha}(x-x')]
+\frac{1}{2}\epsilon_{\mu\nu\alpha\beta}\partial_{\mu}\lambda_{\nu}=0\;.
\label{diff}
\end{equation}
This appears to be very different than the remaining set of Maxwell equations: $\partial_{\nu}F_{\mu\nu}=0$. Maxwell's equations can be obtained by differentiating (\ref{diff}) with respect to $x_{\beta}$. However (\ref{diff}) cannot be expected to be equivalent to Maxwell's theory, since we need to differentiate (\ref{diff}) to obtain Maxwell's equations. For example, it is well known that one can obtain wave equations for the electric and magnetic field by differentiating Maxwell's equations. However, not all solutions to the wave equations will satisfy Maxwell's equations. The solution space of these wave equations is larger than that of Maxwell's equations.
The Hamiltonian formalism is not investigated in this paper.

In \cite{Man1} Mandelstam develops a quantum theory of the electromagnetic field coupled to a complex scalar field without potentials.
He considers the electromagnetic field coupled to a complex scalar field and takes the
gauge invariant variables $\Phi(x,P)$, $\Phi^{*}(x,P)$ and $F_{\mu\nu}(x)$ are taken as the field variables where
\begin{equation}
\Phi(x,P)=\phi(x)\exp\left\{-ie\int_{-\infty}^xd\xi_{\mu} A_{\mu}(\xi)\right\}
\end{equation}
and the integral is taken over the spacelike path $P$. The Lagrangian is given by
\begin{equation}
L=-\partial_{\mu}\Phi^{*}\partial_{\mu}\Phi-m^2\Phi^{*}\Phi-\frac{1}{4}\left\{F_{\mu\nu}\right\}^2
\end{equation}
where $\partial_{\mu}\Phi$ is the gauge invariant derivative of $\Phi$. This approach differs significantly from the approach taken in this paper.
\section{The Hamiltonian Formulation}
In this section I will consider the free electromagnetic field (i.e. $\rho=0$ and $\vec{j}=0$).
The Hamiltonian for the dynamical equations will be derived first and the constraint equations will then be included.
The canonical momenta are given by
\begin{equation}
\vec{\Pi}_E=\frac{\partial L_D}{\partial\dot{\vec{E}}}=-\vec{B}
\end{equation}
and
\begin{equation}
\vec{\Pi}_B=\frac{\partial L_D}{\partial\dot{\vec{B}}}=\vec{E}.
\end{equation}
We therefore have the primary constraints
\begin{equation}
\vec{\phi}_1=\vec{\Pi}_E+\vec{B}\approx 0 \hskip 0.4in and \hskip 0.4in \vec{\phi}_2=\vec{\Pi}_B-\vec{E}\approx 0\;,
\end{equation}
where $\approx$ denotes a weak equality which can only be imposed after the Poisson brackets have been evaluated.  These constraints satisfy
\begin{equation}
\{\phi_{1n}(\vec{x}),\phi_{1m}(\vec{y})\}=\{\phi_{2n}(\vec{x}),\phi_{2m}(\vec{y})\}=0
\end{equation}
and
\begin{equation}
\{\phi_{1n}(\vec{x}),\phi_{2m}(\vec{y})\}=2\delta_{mn}\delta^{3}(\vec{x}-\vec{y})\;.
\end{equation}
Thus, these constraints are second class.

The canonical Hamiltonian $H_C=\int d^3x[\vec{\Pi}_E\cdot\dot{\vec{E}}+\vec{\Pi}_B\cdot\dot{\vec{B}}-L_D]$ is given by
\begin{equation}
H_C=\int d^3x\left[\vec{\phi}_1\cdot\dot{\vec{E}}+\vec{\phi}_2\cdot\dot{\vec{B}}-\vec{E}\cdot\left(\vec{\nabla}\times\vec{E}\right)-
\vec{B}\cdot\left(\vec{\nabla}\times\vec{B}\right)\right]
\end{equation}
and the total Hamiltonian is given by
\begin{equation}
H_T=-\int d^3x\left[\vec{E}\cdot\left(\vec{\nabla}\times\vec{E}\right)+
\vec{B}\cdot\left(\vec{\nabla}\times\vec{B}\right)-\vec{u}_1\cdot\vec{\phi}_1-\vec{u}_2\cdot\vec{\phi}_2\right]\;,
\label{total}
\end{equation}
where $\vec{u}_1$ and $\vec{u}_2$ are undetermined parameters.

For consistency we require that
\begin{equation}
\dot{\vec{\phi}_1}=\{\vec{\phi}_1,H_T\}\approx 0\hskip 0.4in and \hskip 0.4in \dot{\vec{\phi}}_2=\{\vec{\phi}_2,H_T\}\approx 0\;.
\end{equation}
These two equations give
\begin{equation}
\vec{u}_1=\vec{\nabla}\times\vec{B}
\end{equation}
and
\begin{equation}
\vec{u}_2=-\vec{\nabla}\times\vec{E}\;.
\end{equation}
Substituting these expressions for $\vec{u}_1$ and $\vec{u}_2$ into (\ref{total}) gives
\begin{equation}
H_T=\int d^3x\left[\vec{\Pi}_E\cdot\left(\vec{\nabla}\times\vec{B}\right)-\vec{\Pi}_B\cdot\left(\vec{\nabla}\times\vec{E}\right)\right]\; .
\end{equation}
It is easy to see that $\dot{\vec{E}}=\{\vec{E},H_T\}=\vec{\nabla}\times\vec{B}$ and $\dot{\vec{B}}=\{\vec{B},H_T\}=-\vec{\nabla}\times\vec{E}$.

The constraints
\begin{equation}
\phi_3=\vec{\nabla}\cdot\vec{E}\approx 0 \hskip 0.4in and \hskip 0.4in \phi_4=\vec{\nabla}\cdot\vec{B}\approx 0
\end{equation}
must also be included. It is well known that these constraints are preserved under time evolution so there are no additional constraints.
It is easy to show that
\begin{equation}
\{\phi_3(x),\vec{\phi}_1(y)\}=\vec{\nabla}_x\delta^3(\vec{x}-\vec{y})\;, \hskip 0.5in
\{\phi_3(x),\vec{\phi}_2(y)\}=0\;,
\end{equation}
and
\begin{equation}
\{\phi_4(x),\vec{\phi}_1(y)\}=0\;, \hskip 0.5in
\{\phi_4(x),\vec{\phi}_2(y)\}=\vec{\nabla}_x\delta^3(\vec{x}-\vec{y})\;.
\end{equation}
the constraints $\{\vec{\chi}_1,\; \vec{\chi_2}\;,\chi_3\;,\chi_4\}$ are, therefore, a set of second class constraints.

Systems involving second class constraints can be treated using the Dirac bracket. Before introducing the Dirac bracket it will be
convenient to relabel the constraints as follows
\begin{equation}
\eta_1=\phi_{1x},\;\;\;\;\;\;\;\eta_2=\phi_{1y},\;\;\;\;\;\;\;\eta_3=\phi_{1z},\;\;\;\;\;\;\;\eta_4=\phi_{2x}
\end{equation}
and
\begin{equation}
\eta_5=\phi_{2y},\;\;\;\;\;\;\;\eta_6=\phi_{2z},\;\;\;\;\;\;\;\eta_7=\phi_{3},\;\;\;\;\;\;\;\eta_8=\phi_{4}
\end{equation}
and to define the matrix
\begin{equation}
\Delta_{m\vec{x},n\vec{y}}=\{\eta_{m}(\vec{x}),\eta_{n}(\vec{y})\}.
\end{equation}
The Dirac bracket is defined as
\begin{equation}
\{A,B\}_D=\{A,B\}-\Sigma_{m,n}\int d^3xd^3y\{A,\eta_{m}(\vec{x})\}C_{m\vec{x},n\vec{y}}\{\eta_{n}(\vec{y}),B\}\;,
\end{equation}
where $C_{m\vec{x};n\vec{y}}$ is the inverse of $\Delta_{m\vec{x},n\vec{y}}$ :
\begin{equation}
\Sigma_{k}\int d^3z\; C_{m\vec{x},k\vec{z}}\Delta_{k\vec{z},n\vec{y}}=\delta_{mn}\delta^3(\vec{x}-\vec{y})\;.
\end{equation}
The non-zero elements of $C_{m\vec{x},n\vec{y}}$ are given by (see Appendix II)

\begin{equation}
C_{1\vec{x},4\vec{y}}=-\frac{1}{2}\delta^3(\vec{x}-\vec{y})+\frac{1}{8\pi |\vec{x}-\vec{y}|^3}-\frac{3(x^1-y^1)^2}{8\pi |\vec{x}-\vec{y}|^5}\;,
\end{equation}
\begin{equation}
C_{2\vec{x},5\vec{y}}=-\frac{1}{2}\delta^3(\vec{x}-\vec{y})+\frac{1}{8\pi |\vec{x}-\vec{y}|^3}-\frac{3(x^2-y^2)^2}{8\pi |\vec{x}-\vec{y}|^5}\;,
\end{equation}
\begin{equation}
C_{3\vec{x},6\vec{y}}=-\frac{1}{2}\delta^3(\vec{x}-\vec{y})+\frac{1}{8\pi |\vec{x}-\vec{y}|^3}-\frac{3(x^3-y^3)^2}{8\pi |\vec{x}-\vec{y}|^5}\;,
\end{equation}
\begin{equation}
C_{1\vec{x},5\vec{y}}=C_{2\vec{x},4\vec{y}}=-\frac{3(x^1-y^1)(x^2-y^2)}{8\pi |\vec{x}-\vec{y}|^5}\;,
\end{equation}
\begin{equation}
C_{1\vec{x},6\vec{y}}=C_{3\vec{x},4\vec{y}}=-\frac{3(x^1-y^1)(x^3-y^3)}{8\pi |\vec{x}-\vec{y}|^5}\;,
\end{equation}
\begin{equation}
C_{2\vec{x},6\vec{y}}=C_{3\vec{x},5\vec{y}}=-\frac{3(x^2-y^2)(x^3-y^3)}{8\pi |\vec{x}-\vec{y}|^5}\;,
\end{equation}
\begin{equation}
C_{1\vec{x},7\vec{y}}=C_{4\vec{x},8\vec{y}}=\frac{(x^1-y^1)}{4\pi|\vec{x}-\vec{y}|^3}\;,
\end{equation}
\begin{equation}
C_{2\vec{x},7\vec{y}}=C_{5\vec{x},8\vec{y}}=\frac{(x^2-y^2)}{4\pi|\vec{x}-\vec{y}|^3}\;,
\end{equation}
\begin{equation}
C_{3\vec{x},7\vec{y}}=C_{6\vec{x},8\vec{y}}=\frac{(x^3-y^3)}{4\pi|\vec{x}-\vec{y}|^3}\;,
\end{equation}
\begin{equation}
C_{7\vec{x},8\vec{y}}=\frac{1}{2\pi |\vec{x}-\vec{y}|}\;.
\end{equation}
plus terms related by antisymmetry: $C_{m\vec{x};n\vec{y}}=-C_{n\vec{y};m\vec{x}}$.
The advantage of using Dirac brackets instead of Poisson brackets is that the constraints can be set to zero strongly.
The Dirac brackets involving $\vec{E}$ and $\vec{B}$ are
\begin{equation}
\{E_i(\vec{x}),E_j(\vec{y})\}_D=\{B_i(\vec{x}),B_j(\vec{y})\}_D=0
\end{equation}
and
\begin{equation}
\{E_i(\vec{x}),\Pi_{Ej}(\vec{y})\}_D=\frac{1}{2}\left[\delta^3(\vec{x}-\vec{y})\delta_{ij}+
\frac{\partial^2}{\partial x^i\partial x^j}\left(\frac{1}{4\pi|\vec{x}-\vec{y}|}\right)\right]
\;.
\end{equation}
It is interesting to note that in the standard approach in the Coulomb gauge one finds \cite{Wei1} that $A_i(\vec{x})$ and $\Pi^j(\vec{x})$ satisfy the same Dirac bracket up to a factor of 1/2.
The other Dirac brackets involving the momenta can be found from these using the constraints (i.e. $\{E_i(\vec{x}),B_{j}(\vec{y})\}_D
=-\{E_i(\vec{x}),\Pi_{Ej}(\vec{y})\}_D$).

The constraints $\vec{\nabla}\cdot\vec{E}=0$ and $\vec{\nabla}\cdot\vec{B}=0$ imply that the longitudinal components of $\vec{E}$ and $\vec{B}$ vanish. The constraints $\vec{\phi}_1=0$ and $\vec{\phi}_2=0$ then imply that the longitudinal components of $\vec{\Pi}_E$ and $\vec{\Pi}_B$ vanish.
The four independent degrees of freedom can therefore be taken to be the transverse components of $\vec{E}$ and $\vec{\Pi}_E$.

One could also have started with $L_D+L_C$ and obtained similar results with two additional constraints, $\Pi_{\lambda_{1}}\approx 0$ and
$\Pi_{\lambda_{2}}\approx 0$.
\section{Conclusion}
In this paper I found a Lagrangian that gives the dynamical Maxwell equations and included the constraint equations using Lagrange multipliers. In the Hamiltonian formalism one finds that the canonical momenta $\vec{\Pi}_E$ and $\vec{\Pi}_B$ are constrained
giving 6 second class constraints at each point in space. Gauss's law and $\vec{\nabla}\cdot\vec{B}=0$ can than be added in as additional constraints giving a total of 8 constraints. This leaves 4 independent degrees of freedom, as expected.
\section*{Acknowledgements}
This research was supported by the  Natural Sciences and Engineering Research
Council of Canada.
\section{Appendix I}
In this appendix I examine the transformation properties of $L=L_D+L_S$ and the constraints under Lorentz transformations and find  covariant expressions for these quantities. For notational simplicity define
\begin{equation}
\vec{a}=\frac{\partial\vec{E}}{\partial t}-\vec{\nabla}\times\vec{B}+4\pi\vec{j}
\end{equation}
and
\begin{equation}
\vec{b}=\frac{\partial\vec{B}}{\partial t}+\vec{\nabla}\times\vec{E}
\end{equation}

Now consider the constraints. They can be written as
\begin{equation}
\vec{\nabla}\cdot\vec{E}-4\pi\rho=\partial_{\alpha}F^{\alpha}_{\;\;0}+4\pi j_0=0
\label{LT}
\end{equation}
and
\begin{equation}
\vec{\nabla}\cdot\vec{B}=\partial_{\alpha}G^{\alpha}_{\;\;0}=0\;,
\end{equation}
where $F_{\mu\nu}$ is the electromagnetic field strength tensor, $G^{\mu\nu}=\frac{1}{2}\epsilon^{\mu\nu\alpha\beta}F_{\alpha\beta}$
is its dual and $j^{\mu}$ is the current density. Consider the constraints in the inertial frame $\bar{S}$. Under the Lorentz transformation $\Lambda_{\mu}^{\;\;\nu}$ from $\bar{S}$ to the inertial frame $S$
\begin{equation}
\bar{\partial}_{\alpha}\bar{F}^{\alpha}_{\;\;0}+4\pi\bar{j}_0=\Lambda_{0}^{\;\;\mu}(\partial_{\alpha}F^{\alpha}_{\;\;\mu}+4\pi j_\mu)=0
\end{equation}
and
\begin{equation}
\bar{\partial}_{\alpha}\bar{G}^{\alpha}_{\;\;0}=\Lambda_{0}^{\;\;\mu}\partial_{\alpha}G^{\alpha}_{\;\;\mu}=0\;.
\end{equation}
The constraints in $S$ are then
\begin{equation}
\Lambda_{0}^{\;\;\mu}(\partial_{\alpha}F^{\alpha}_{\;\;\mu}+4\pi j_\mu)=0
\;\;\;\;\;\;\;\;\;\;\; and \;\;\;\;\;\;\;\;\;\;\;
\Lambda_{0}^{\;\;\mu}\partial_{\alpha}G^{\alpha}_{\;\;\mu}=0\;.
\label{Constraints}
\end{equation}
They can also be written as
\begin{equation}
\vec{\nabla}\cdot\vec{E}-4\pi\rho+\vec{\Lambda}\cdot\vec{a}=0
\label{LTC1}
\end{equation}
and
\begin{equation}
\vec{\nabla}\cdot\vec{B}+\vec{\Lambda}\cdot\vec{b}=0\;,
\label{LTC2}
\end{equation}
where $\vec{\Lambda}_k=\frac{\Lambda_{0}^{\;\;k}}{\Lambda_{0}^{\;\;0}}$.

The Lagrangian $L=L_D+L_S$ can be written as
\begin{equation}
L=F_{0\mu}\partial_{\alpha}G^{\mu\alpha}-G_{0\mu}\partial_{\alpha}F^{\mu\alpha}+8\pi G_{0\mu}j^{\mu}\;.
\label{LTL}
\end{equation}
Consider the Lagrangian $\bar{L}=\bar{L}_d+\bar{L}_S$ in the inertial frame $\bar{S}$. Under the Lorentz transformation $\Lambda_{\mu}^{\;\;\nu}$ from $\bar{S}$ to the inertial frame $S$
\begin{equation}
\bar{F}_{0\mu}\bar{\partial}_{\alpha}\bar{G}^{\mu\alpha}-\bar{G}_{0\mu}\partial_{\alpha}\bar{F}^{\mu\alpha}+8\pi \bar{G}_{0\mu}\bar{j}^{\mu}=
\Lambda_{0}^{\;\;\nu}(F_{\nu\mu}\partial_{\alpha}G^{\mu\alpha}-G_{\nu\mu}\partial_{\alpha}F^{\mu\alpha}+8\pi G_{\nu\mu}j^{\mu})\;.
\label{LHS}
\end{equation}
The Lagrangian in $S$ is then given by.
\begin{equation}
L=\Lambda_{0}^{\;\;\mu}(F_{\mu\nu}\partial_{\alpha}G^{\nu\alpha}-G_{\mu\nu}\partial_{\alpha}F^{\nu\alpha}+8\pi G_{\mu\nu}j^{\nu})\;.
\label{Rel}
\end{equation}
Expressing this Lagrangian in terms of $\vec{E}$ and $\vec{B}$ gives
\begin{equation}
L=\Lambda_{0}^{\;\;0}\left\{\vec{b}\cdot\vec{E}-\left(\vec{a}+4\pi\vec{j}\right)\cdot\vec{B}+
\vec{\Lambda}\cdot\left[\vec{E}(\vec{\nabla}\cdot\vec{B})
-\vec{B}(\vec{\nabla}\cdot\vec{E})
+\vec{B}\times\vec{b}+
\vec{E}\times\left(\vec{a}+4\pi\vec{j}\right)+
8\pi\rho\vec{B}\right]\right\}.
\label{LT1}
\end{equation}
The field equations, using the constraints (\ref{LTC1}) and (\ref{LTC2}),
generated by this Lagrangian are
\begin{equation}
\vec{a}+\vec{\Lambda}\times\vec{b}-(\vec{a}\cdot\vec{\Lambda})\vec{\Lambda}=0
\label{LT2}
\end{equation}
and
\begin{equation}
\vec{b}-\vec{\Lambda}\times\vec{a}-(\vec{b}\cdot\vec{\Lambda})\vec{\Lambda}=0\;.
\label{LT3}
\end{equation}
Taking the inner product of both sides of these equations with $\vec{\Lambda}$ gives
\begin{equation}
(1-\Lambda^2)(\vec{\Lambda}\cdot\vec{a})=0\;\;\;\;\;\;\;\;\;\;\; and \;\;\;\;\;\;\;\;\;\;\;(1-\Lambda^2)(\vec{\Lambda}\cdot\vec{b})=0\;,
\end{equation}
where $\Lambda^2=\vec{\Lambda}\cdot\vec{\Lambda}$. Thus, for $\Lambda^2\neq 1$, which will be the case for Lorentz transformations, we have
$\vec{\Lambda}\cdot\vec{a}=\vec{\Lambda}\cdot\vec{b}=0$
Substituting (\ref{LT3}) into (\ref{LT2}) and using $\vec{\Lambda}\cdot\vec{a}=\vec{\Lambda}\cdot\vec{b}=0$ gives
\begin{equation}
\vec{a}=\frac{\partial\vec{E}}{\partial t}-\vec{\nabla}\times\vec{B}+4\pi\vec{j}=0\;.
\end{equation}
A similar argument gives
\begin{equation}
\vec{b}=\frac{\partial\vec{B}}{\partial t}+\vec{\nabla}\times\vec{E}=0\;.
\end{equation}

In the above derivation of Maxwell's equations I used $|\vec{\Lambda}|\neq 1$, which is equivalent to
$-(\Lambda_{0}^{\;\;0})^2+\Sigma_k\Lambda_{0}^{\;\;k}\Lambda_{0}^{\;\;k}\neq 0$, and $\Lambda_{0}^{\;\;0}\neq 0$, both of which are true for Lorentz transformations.
It is easy to see that the Lagrangian and constraints will still produce Maxwell's equations if $\Lambda_{0}^{\;\;\mu}$ in (\ref{Constraints}) and (\ref{Rel}) is replaced by any constant four-vector $\alpha^{\mu}$ satisfying $\alpha^{\mu}\alpha_{\mu}\neq 0$ and $\alpha^0\neq 0$. The condition $\alpha^{\mu}\alpha_{\mu}\neq 0$ constrains $\alpha^{\mu}$ to be non-null. If $\alpha^{\mu}$ is a space-like four vector there exists an inertial frame in which $\alpha^0=0$. Therefore, if the Lagrangian and constraints are to produce Maxwell's theory in all inertial frames of reference $\alpha^{\mu}$ must be a time-like four-vector.
Thus, the Lagrangian
\begin{equation}
L_{\alpha}=\alpha^{\mu}\left[F_{\mu\nu}\partial_{\beta}G^{\nu\beta}-G_{\mu\nu}\partial_{\beta}F^{\nu\beta}
+8\pi G_{\mu\beta}j^{\beta}\right]
\end{equation}
and the constraints
\begin{equation}
\alpha^{\mu}[\partial_{\beta}F^{\beta}_{\;\;\mu}+4\pi j_\mu]=0\;\;\;\;\;\;\;\;\;\;\; and \;\;\;\;\;\;\;\;\;\;\;
\alpha^{\mu}[\partial_{\beta}G^{\beta}_{\;\;\mu}]=0
\end{equation}
are Lorentz scalars and give Maxwell equations for any constant time-like four-vector $\alpha^{\mu}$. For simplicity I have imposed the constraints outside the variational principle so that the Lagrange multipliers do not appear in the equations for the fields.
\section{Appendix II}
In this appendix I derive a few of the $C_{i\vec{x},j\vec{y}}$ to illustrate how the calculations are done.
I will assume that $C_{n\vec{x},n\vec{y}}=0$ (no sum on $n$) and that $C_{m\vec{x},n\vec{y}}$ is a function of $(x^1-y^1,x^2-y^2,x^3-y^3)$.
These assumptions will be justified by the success of the calculations. Some of the equations for $C_{i\vec{x},j\vec{y}}$ are differential equations, so boundary conditions need to be imposed. I will take these conditions to be
\begin{equation}
C_{i\vec{x},j\vec{y}}\rightarrow 0 \hskip .3in as\hskip .3in |\vec{x}-\vec{y}|\rightarrow\infty\;.
\label{BC}
\end{equation}
Define $P_{i\vec{x},j\vec{y}}$ by
\begin{equation}
P_{m\vec{x},n\vec{y}}=\Sigma_{k}\int d^3z\; C_{m\vec{x},k\vec{z}}\Delta_{k\vec{z},n\vec{y}}\;.
\end{equation}
Consider $m=7$ and $n=1$
\begin{equation}
P_{7\vec{x},1\vec{y}}=C_{7\vec{x},4\vec{y}}=0
\end{equation}
and $m=1$ and $n=4$
\begin{equation}
P_{1\vec{x},4\vec{y}}=-\frac{\partial}{\partial y^1} C_{1\vec{x},8\vec{y}}=0\;.
\end{equation}
This implies that $C_{1\vec{x},8\vec{y}}$ is independent of $x^1-y^1$. The boundary condition (\ref{BC}) then gives
$C_{1\vec{x},8\vec{y}}=0$.

Now consider
\begin{equation}
P_{1\vec{x},1\vec{y}}=-2C_{1\vec{x},4\vec{y}}-\frac{\partial}{\partial y^1}C_{1\vec{x},7\vec{y}}=\delta^3
(\vec{x}-\vec{y})\;,
\label{1}
\end{equation}
\begin{equation}
P_{1\vec{x},2\vec{y}}=-2C_{1\vec{x},5\vec{y}}-\frac{\partial}{\partial y^2}C_{1\vec{x},7\vec{y}}=0\;,
\label{2}
\end{equation}
\begin{equation}
P_{1\vec{x},3\vec{y}}=-2C_{1\vec{x},6\vec{y}}-\frac{\partial}{\partial y^3}C_{1\vec{x},7\vec{y}}=0\;,
\label{3}
\end{equation}
and
\begin{equation}
P_{1\vec{x},8\vec{y}}=-\frac{\partial}{\partial y^1}C_{1\vec{x},4\vec{y}}-\frac{\partial}{\partial y^2}C_{1\vec{x},5\vec{y}}
-\frac{\partial}{\partial y^3}C_{1\vec{x},6\vec{y}}=0\;.
\label{4}
\end{equation}
Substituting (\ref{1}), (\ref{2}) and (\ref{3}) into (\ref{4}) gives
\begin{equation}
\nabla^2_yC_{1\vec{x},7\vec{y}}=-\frac{\partial}{\partial y^1}\delta^3(\vec{x}-\vec{y})\;.
\end{equation}
This equation has the solution
\begin{equation}
C_{1\vec{x},7\vec{y}}=\frac{(x^1-y^1)}{4\pi|\vec{x}-\vec{y}|^3}\;.
\end{equation}
Equations (\ref{2}) and (\ref{3}) now give
\begin{equation}
C_{1\vec{x},5\vec{y}}=-\frac{3(x^1-y^1)(x^2-y^2)}{8\pi |\vec{x}-\vec{y}|^5}
\end{equation}
and
\begin{equation}
C_{1\vec{x},6\vec{y}}=-\frac{3(x^1-y^1)(x^3-y^3)}{8\pi |\vec{x}-\vec{y}|^5}\;.
\end{equation}


\begin{thebibliography}{99}
\bibitem{Jac1}
J.D. Jackson, Classical electrodynamics (3rd edition, Wiley, 1999).
\bibitem{Lan1}
L.D. Landau and E.M. Lifshitz, The classical theory of fields (Pergamon, Oxford, 1962).
\bibitem{Dir1}
P.A.M. Dirac, \textit{Lectures on Quantum Mechanics} (Dover Publications, Mineola, New York, 1964), pp 1-24.
\bibitem{Dir2}
P.A.M. Dirac, Can. J. Math. 2, 129 (1950).
\bibitem{Inf1}
L. Infeld and J. Plebanski, Proc. Roy. Soc. A222, 224 (1954)
\bibitem{Bor1}
M. Born, Proc. Roy. Soc. A143, 410 (1934)
\bibitem{Bor2}
M. Born and L. Infeld, Proc. Roy. Soc. A144, 425 (1934)
\bibitem{Fiu1}
J. Fiutak and M. Zukowski, J. Phys. A14, 3229 (1981)
\bibitem{Ogi1}
V. I. Ogievetskii and I. V. Polubarinov, Zh. Eksp. Teor. Fiz. 43, 1365 (1962), JETP 16, 969 (1963)
\bibitem{Str1}
V. I. Strazhev and P. L. Shkol'nikov, Sov. Phys. J. 32, 378 (1989)
\bibitem{Man1}
S. Mandelstam, Annals Phys. 19, 1 (1962)
\bibitem{Wei1}
Steven Weinberg, \textit{The Quantum Theory of Fields: Volume I Foundations} (Cambridge University Press 1995), p 348.
\end{thebibliography}
\end{document}